\documentstyle[preprint,aps,prbbib]{revtex}
\begin{document}
\begin{center}{\Large Is Galaxy Dark Matter a Property of Spacetime?}\end{center}
\vspace{2 mm}
\begin{center}{Moshe Carmeli}\end{center}
\begin{center}{Department of Physics, Ben Gurion University, Beer Sheva 84105, Israel} 
\end{center}
\begin{abstract}
We describe the motion of a particle in a central field in an expanding universe. 
Use is made of a double expansion in $1/c$ and $1/\tau$, where
$c$ and $\tau$ are the speed of light and the Hubble time. In the lowest approximation
the rotational velocity is shown to satisfy $v^4=\frac{2}{3}GMcH_0$, where $G$ 
is Newton's gravitational
constant, $M$ is the mass of the central body (galaxy) and $H_0$ is the Hubble 
constant. This formula satisfies observations of stars moving in spiral and elliptical 
galaxies, and in accordance with the familiar Tully-Fisher law.
\end{abstract}
\vspace{8mm}
\begin{center}{PACS numbers: 04.25.Nx}\end{center} 
\newpage
\section {Introduction} The problem of motion in general relativity theory is 
as old as general relativity itself. Soon after the theory was published, Einstein 
and Grommer~\cite{E-G} showed that the equations of motion in general relativity
follow from the Einstein feild equations rather that have to be postulated independently
as in other theories. This is a consequence of the nonlinearity of the field
equations and the Bianchi identities. Much work was done since 
then~\cite{E-I-H,E-I,F,F-TS,I,I-P,I-S,B-P,C-PL9,C-PL11,C-AP30,C-PR138,C-NC,C-AP34,C-AP35,C-PR140}
and the problem of motion in the gravitational field of an isolated system
seems to be well understood these days~\cite{D}.

We here formulate the problem of motion in an expanding universe, a topic which
has not been discussed so far. The problem is of considerable importance in astronomy
since stars moving in spiral and elliptical galaxies show serious deviation 
from Newtonian gravity and as is well known, the latter follows from general
relativity theory in a certain approximation~\cite{C-CF}. It follows that the Hubble
expansion imposes an extra constraint on the motion; the usual assumptions
made in deriving Newtonian gravity from general relativity are not sufficient
in an expanding universe. The star is not isolated from the "flow" of matter
in the universe. When this is taken into account, along with Newton's gravity,
the result is a motion which satisfies a different law from the one determining
the planetary motion in the solar system.
\section{Geodesic Equation}
The equation that describes the motion of a simple particle is the geodesic
equation. It is a direct result of the Einstein field equations $G_{\mu\nu}=
\kappa T_{\mu\nu} \left(\kappa=8\pi G/c^4\right)$. The restricted Bianchi identities
$\nabla_{\nu}G^{\mu\nu}\equiv 0$ implies the covariant conservation law $\nabla_{\nu}T^{\mu\nu}=0$.   
When volume-integrated, the latter yields the geodesic equation. To obtain the 
Newtonian gravity it is sufficient to assume the approximate forms for the
metric: $g_{00}=1+\frac{2\phi}{c^2}$, $g_{0k}=0$ and $g_{kl}=-\delta_{kl}$, 
where $k,l=1,2,3$, and $\phi$ 
a function that is determined by the Einstein field equations. In the lowest 
approximation in $1/c$ one then has $$\frac{d^2x^k}{dt^2}=-\frac{\partial\phi}{\partial x^k} \eqno(1)$$
$$\nabla^2\phi=4\pi G\rho \eqno(2)$$ where $\rho$ is the mass density. For a
central body $M$ one then has $\phi=-GM/R$ and Eq. (2) yields, for circular 
motion, the first integral $$v^2=GM/R \eqno(3)$$ where $v$ is the rotational
velocity of the particle.
\section {Hubble's Law} The Hubble law asserts that faraway galaxies recede 
from each other at velocities proportional to their relative distances,
${\bf v}=H_0{\bf R}$, with ${\bf R}=\left(x,y,z\right)$. $H_0$ is the universal 
proportionality constant (at each cosmic time). Obviously the Hubble law can
be written as $\left(\tau=H_0^{-1}\right)$ $$\tau^2v^2-\left(x^2+y^2+z^2\right)=0
\eqno(4)$$ and thus, when gravity is negligible, cosmology can be formulated
as a new special relativity with a new Lorentz-like 
transformation~\cite{C-FP25,C-CTP4a,C-CTP4b,C-FP,C-CTP5}.
Gravitation, however, does not permit global linear relations like Eq. (4) and
the latter has to be adopted to curved space. To this end one has to modify Eq. (4)
to the differential form and to adjust it to curved space. The generalization of 
Eq. (4) is, accordingly, $$ds^2=g'_{\mu\nu}dx^{\mu}dx^{\nu}=0 \eqno(5)$$ with 
$x^0=\tau v$. Since the universe expands radially (it is assumed to be homogeneous 
and isotropic), it is convenient to use spherical coordinates $x^k=\left(R,\theta,\phi\right)$ 
and thus $d\theta=d\phi=0$. We are still entitled to adopt coordinate conditions,
which we choose as $g'_{0k}=0$ and $g'_{11}=g'^{-1}_{00}$. Equation (5) reduces to
$$\frac{dR}{dv}=\tau g'_{00} \eqno(6)$$ This is Hubble's law taking into account
gravitation, and hence dilation and curvature. When gravity is negligible,
$g'_{00}\approx 1$ thus $\frac{dR}{dv}=\tau$ and by integration, $R=\tau v$ or
$v=H_0 R$ when the initial conditions are chosen appropriately.
\section {Phase Space} As is seen, the Hubble expansion causes constraints on
the structure of the universe which is expressed in the phase space of distances 
and velocities, exactly the observables. The question arises: What field equations
the metric tensor $g'_{\mu\nu}$ satisfies? We {\it postulate} that $g'_{\mu\nu}$ 
satisfies the Einstein field equations in the phase space, $G'_{\mu\nu}=KT'_{\mu\nu}$,
with $K=\frac{8\pi k}{\tau^4}$, and $k=\frac{G\tau^2}{c^2}$. Accordingly, in cosmology
one has to work in both the real space and in the phase space. Particles follow
geodesics of both spaces (in both cases they are consequences of the Bianchi 
identities). For a sperical solution in the phase space, similarly to the
situation in the real space, we have in the lowest approximation in $1/\tau$
the following: $g'_{00}=1+\frac{2\psi}{\tau^2}$, $g'_{0k}=0$ and $g'{kl}=-\delta_{kl}$,
with $\nabla^2\psi=4\pi k\rho$. For a spherical solution we have $\psi=-kM/R$ 
and the geodesic equation yields $$\frac{d^2x^k}{dv^2}=-\frac{\partial\psi}{\partial x^k} 
\eqno(7)$$ with the first integral $$\left(\frac{dR}{dv}\right)^2=\frac{kM}{R} \eqno(8)$$
for a rotational motion. Integration of Eq. (8) then gives 
$$R=\left(\frac{3}{2}\right)^{2/3}\left(kM\right)^{1/3}v^{2/3} \eqno(9)$$ Inserting
this value of $R$ in Eq (3) we obtain $$v^4=\frac{2}{3}GMcH_0 \eqno(10)$$   
\section {Galaxy dark matter} The equation of motion (10) has a direct relevance
to the problem of the existance of the galaxy dark matter. As is well known, 
observations show that the fourth power of the rotational velocity of stars in 
some galaxies is proportional to the luminousity of the galaxy (Tully-Fisher
law), $v^4\propto L$. Since the luminousity, by turn, is proportional to the 
mass $M$ of the galaxy, $L\propto M$, it follows that $v^4\propto M$, independent 
of the radial distance of the star from the center of the galaxy, and in violation 
to Newtonian gravity. Here came the idea of galaxy dark matter or, alternatively,
modification of Newton's gravity in an expanding universe. 
\vspace {8mm}
\newline
In this paper we have seen how a careful application of general relativity
theory gives an answer to the problem of motion of stars in galaxies in an
expanding universe. If Einstein's general relativity theory is valid, then it
appears that the galaxy halo dark matter is a property of spacetime and not 
some physical material. The situation resembles that existed at the beginning
of the century with respect to the problem of the advance of the perihelion of
the planet Mercury which general relativity theory showed that it was a property
of spacetime (curvature). 
\section{Acknowlegements} It is a pleasure to thank Y. Ne'eman for many discussions,
illuminating remarks and much encouragements. Thanks are also due to G. Erez, 
B. Carr, O. Lahav and N. van den Bergh for useful remarks.

\end{document}